\documentstyle[12pt,bezier]{article}
\def\e{\mbox{e}}

\begin{document}
\title{Semiclassical approach for multiparticle production in scalar
theories}
\author{D.~T.~Son\\
{\small\em  Institute for Nuclear Research of the Russian Academy of
Sciences}\\
{\small\em 60th October Anniversary Prospect 7a, Moscow 117312 Russia}\\
and\\
{\small\em Department of Physics and Astronomy, Rutgers University}\\
{\small\em Piscataway, New Jersey 08855--0849 USA}}
\date{May 1995}
\maketitle
\begin{abstract}
We propose a semiclassical approach to calculate multiparticle cross
sections in scalar theories, which have been strongly argued to have
the exponential form $\exp(\lambda^{-1}F(\lambda n,\epsilon))$ in the
regime $\lambda\to0$, $\lambda n$, $\epsilon=$ fixed, where $\lambda$
is the scalar coupling, $n$ is the number of produced particles, and
$\epsilon$ is the kinetic energy per final particle.  The formalism is
based on singular solutions to the field equation, which satisfy
certain boundary and extremizing conditions.  At low multiplicities
and small kinetic energies per final particle we reproduce in the
framework of this formalism the main perturbative results.  We also
obtain a lower bound on the tree--level cross section in the
ultra--relativistic regime.
\end{abstract}
\vskip .5in
{RU--95--31\hfill}
\newpage

\section{Introduction}

Recently, the problem of multiparticle production in weakly coupled
scalar field theories has received close attention.  This problem has
been initiated by the qualitative observation \cite{Cornwall,Goldberg}
that in the ${\lambda\over4}\phi^4$ theory at tree level the
probability of processes producing large number of bosons exhibit
factorial dependence on the multiplicity of the final state.  This
dependence originates from the large number of tree graphs
contributing to the process of multiparticle production: at
multiplicity $n\gg1$ the number of graphs is of order $n!$.  At
$n\sim1/\lambda$ this factor is sufficient to compensate the
suppression due to the smallness of the coupling constant, and the
tree--level multiparticle cross sections become large.  Much efforts
have been made to understand how this behavior is changed by loop
corrections, though a conclusive result on this issue is still lacking
right now.

So far, quantitative calculations have been performed mostly at, or
near, the multiparticle threshold, where some perturbative techniques
have been developed and extensively explored.  The tree amplitude of
transition from one initial virtual particle to $n$ real bosons at
rest (the $1\to n$ process) can be computed either by summing Feynman
diagrams or by using some appropriate classical solution
\cite{Vol,AKP,Brown} and the result reads
\begin{equation}
  A^{tree}_n(0)=n!\left({\lambda\over8}\right)^{n-1\over2}
  \label{Atree}
\end{equation}
(the boson mass in this formula and further is set to 1).  The same
methods have been applied for calculating the amplitude beyond the
threshold or beyond the tree level \cite{LRST}.  In the first case it
has been found that when the final particles are non--relativistic,
the tree amplitude is an exponent of the total kinetic energy of final
particles,
\begin{equation}
  A^{tree}_n(\epsilon)=A^{tree}_n(0)
  \e^{-{5\over6}n\epsilon}
  \label{intro_1}
\end{equation}
where $A^{tree}_n(0)$ is given by eq.(\ref{Atree}) and $\epsilon$ is
the kinetic energy per particle in the final state.  The exponential
fall of the tree amplitude beyond threshold in eq.(\ref{intro_1}) is
not sufficient, however, to make the cross section small.  The second
result concerns loop corrections at exact threshold and reads that the
leading--$n$ contributions from each loop level (namely, the $\lambda
n^2$ contribution from the first loop, $\lambda^2n^4$ from the second
and $\lambda^kn^{2k}$ from the $k$--th) sum up to an exponent, so at
not very large $n$, (when subleading on $n$ contributions can be
neglected, presumably at $\lambda n\ll1$) the $1\to n$ amplitude at
threshold has the form,
\begin{equation}
  A_n(0)=A^{tree}_n(0)\e^{B\lambda n^2}
  \label{intro_2}
\end{equation}
where $B$ is a constant that depends on the number of spatial dimensions,
\[
  B=\int\!{d{\bf k}\over(2\pi)^d}\,{9\over8\omega_{\bf k}
  (\omega_{\bf k}^2-1)(\omega_{\bf k}^2-4)},\qquad
  \omega_{\bf k}=\sqrt{{\bf k}^2+1}
\]
In particular, in (3+1) dimensions ($d=3$), the numeric value of $B$
is
\[
  B=-{1\over64\pi^2}(\ln(7+4\sqrt{3})-i\pi)
\]

The physically interesting quantity is however not the amplitude, but
the cross section, or transition rate.  For the $1\to n$ process near
threshold this quantity is easy to evaluate at $n\ll1/\lambda$, having
on hand the two results above.  In fact, if $\epsilon$ is small enough
for the amplitude to be constant in the whole phase volume, the cross
section is equal to
\[
  \sigma(E,n)=|A_n(\epsilon)|^2V_n
\]
where $A_n(\epsilon)=A^{tree}_n\exp(-{5\over6}n\epsilon+B\lambda
n^2)$, and $V_n$ is the bosonic phase volume.  An important point to
note is that at large $n$, $V_n$ has the exponential form except from
the factor of $1/n!$,
\begin{equation}
  V_n\propto{1\over n!}\exp\left({dn\over2}\left(
  \ln{\epsilon\over\pi d}+1\right)+{d-2\over4}n\epsilon+O(n\epsilon^2)
  \right)
  \label{Vn}
\end{equation}
and now it is easy to verify that the cross section is exponential,
\begin{equation}
  \sigma(E,n)\propto\exp\left(n\ln{\lambda n\over16}-n+
  {dn\over2}\left(\ln{\epsilon\over\pi d}+1\right)+
  \left({d-2\over4}-{5\over3}\right)n\epsilon+2\mbox{Re}B\lambda^2n
  \right)
  \label{page3}
\end{equation}
Though eq.(\ref{page3}) is valid only at small $\epsilon$ and $\lambda
n$, the form of $\sigma(E,n)$ strongly supports the hypothesis that in
the most interesting regime $\lambda n\sim1$, $\epsilon\sim1$ the
cross section is also exponential \cite{LRST}.
\begin{equation}
  \sigma(E,n)\propto\exp\left({1\over\lambda}F(\lambda n,\epsilon)
  \right),\qquad\epsilon={E-n\over n}
  \label{intro_exp}
\end{equation}
Moreover, there are indications \cite{LST} that the exponent
$F(\lambda n,\epsilon)$ is independent of the few--particle initial
state (i.e. the cross section of $2\to n$, $3\to n$, etc. processes
coincide, with exponential accuracy, with that of $1\to n$). The
function $F(\lambda n,\epsilon)$ is unknown, but some terms of its
expansion at small $\lambda n$ and $\epsilon$ can be found from
(\ref{page3}),
\[
  F(\lambda n,\epsilon)=\lambda n\ln{\lambda n\over16}-\lambda n
  +{d\over2}\left(\ln{\epsilon\over\pi d}+1\right)\lambda n+
  \left({d-2\over4}-{5\over3}\right)\lambda n\epsilon+
  2\mbox{Re}B\lambda^2n^2+
\]
\begin{equation}
  +O(\lambda^3n^3)+O(\lambda^2n^2\epsilon)+
  O(\lambda n\epsilon^2)
  \label{Fpert}
\end{equation}

The situation that has emerged here shows a complete analogy to the
instanton--like processes at high energies.  As in the latter case,
the exponential form of the multiparticle cross sections is a strong
argument in favor to the semiclassical calculability of the exponent
$F$, but says nothing about the nature of possible calculation
schemes.

In this paper we propose a semiclassical method to calculate the
multiparticle cross section.  By using the coherent state formalism,
we reduce the calculation to the problem of solving classical field
equation with certain boundary conditions in the asymptotic regions
$t\to\pm\infty$.  The technique, as well as the boundary value problem
are very similar to the those in the case of instanton transitions.
The most essential difference from the latter case is that the
boundary value problem for multiparticle production possesses only
singular solutions.  In particular, the field configuration defining
the cross section is the one that is singular at one point $t={\bf
x}=0$ and regular elsewhere in the Minkowskian space--time.  Note that
some other approaches utilizing singular solutions have been proposed
recently for both multiparticle processes in scalar theories and
instanton--like transitions
\cite{VolLandau,Khlebnikov,CornTikt,DiakPetr}, most being inspired by
the Landau procedure for calculating the semiclassical matrix elements
\cite{Landau}.  We emphasize, however, that in the framework of our
formalism the singular field configuration is determined in a unique
way by the boundary conditions and the structure of its singularity in
the Minkowskian space--time.

The field configuration with the required properties can also be found
in a different setting.  Namely, if one makes analytical continuation
to the complex times and looks for solutions to the boundary value
problem which is singular on some surface (in the simplest case the
surface lies in the Euclidean space--time) and extremizes the
transition rate over all possible forms of this surface , one obtains
the same field configuration as one would find by solving the boundary
value problem.  Actually, this formalism sometimes appears to be
simpler and will be applied for making quantitative calculations in
this paper, which include reproducing eq.(\ref{page3}) at small
$\lambda n$ and $\epsilon$ and finding a lower bound on the
tree--level cross section in the ultra--relativistic limit of the
final state.

This paper is organized as follows.  In Sect.2 we derive the classical
problem for the multiparticle cross sections.  Sect.3 is devoted to
the tree amplitude at threshold.  A simplified, entirely Euclidean
version of the classical problem is presented and the perturbative
result for the energy dependence of the tree amplitude near threshold
is reproduced.  In the opposite, ultra--relativistic, limit, we obtain
a lower bound on the tree cross section.  In Sect. 4 we consider the
amplitude (with loops) at exact threshold, where the procedure for
calculating the amplitude is derived from the general formalism.  We
reproduce the exponentiation factor coming from leading--$n$ loop in
the limit $\lambda n\ll1$.  Finally, Sect. 5 contains concluding
remarks.

\section{General formalism}

\subsection{The boundary value problem}

We consider the scalar field theory without symmetry breaking in
$(d+1)$--dimensional space--time,
\[
  S=\int\!d^{d+1}x\,\left({1\over2}(\partial_{\mu}\phi)^2-
  {1\over2}\phi^2-{\lambda\over4}\phi^4\right)
\]
The quantity in interest is the total transition rate from an
initial few--particle state to all possible final states having given
energy $E$ and multiplicity $n$.  One writes,
\begin{equation}
  \sigma(E,n)=\sum_f |\langle f|P_EP_n\hat{S}\hat{A}|0\rangle|^2
  \label{sigma}
\end{equation}
where $\hat{A}$ is the operator that creates the initial state from
the vacuum, $\hat{S}$ is the $S$--matrix, $P_E$ and $P_n$ are the
projection operators to states with energy $E$ and number of particles
$n$, and the sum runs over all final states $|f\rangle$.  Different
choices of the operator $\hat{A}$ corresponds to different initial
states: for example, $\hat{A}=\phi$ corresponds to the $1\to n$
process (for the operator $\hat{A}$ describing the $2\to n$ process
see \cite{LST}).  We recall perturbative calculations in \cite{LST}
indicating that $F$ does not depend on the particular choice of
$\hat{A}$, providing the latter is independent of $\lambda$
parametrically.  Making use of this fact, we will calculate
(\ref{sigma}) for the operator $\hat{A}$ most convenient for our
purpose.  Namely, we choose $\hat{A}$ in the following exponential
form,
\[
  \hat{A}=\e^{j\phi(0)}
\]
where $j$ is some arbitrary number.

Following the technique of \cite{RST}, we derive the classical
boundary value problem for $\sigma(E,n)$.  Using the coherent state
formalism \cite{KRT}, one rewrites eq.(\ref{sigma}) in the following
integral form,
\[
  \sigma(E,n)=\int\!db_{\bf k}^*\,db_{\bf k}\,d\xi\,d\eta\,
  {\cal D}\phi\,{\cal D}\phi'\,\exp\left(
  -\int\!d{\bf k}\,b_{\bf k}^*b_{\bf k}\e^{i\omega_k\xi+i\eta}
  +iE\xi+in\eta+\right.
\]
\begin{equation}
  +\left.B_i(0,\phi_i)+B_f(b_{\bf k}^*,\phi_f)+B_i^*(0,\phi'_i)+
  B_f^*(b_{\bf k},\phi'_f)+iS[\phi]-iS[\phi']
  +j\phi(0)+j\phi'(0)\right)
  \label{int_repr}
\end{equation}
In eq.(\ref{int_repr}), $B$'s stay for the boundary terms,
\[
  B_i(0,\phi_i)=-{1\over2}\int\!d{\bf k}\,\omega_{\bf k}\phi_i({\bf k})
                \phi_i(-{\bf k})
\]
\[
  B_f(b^*_{\bf k},\phi_f)=-{1\over2}\int\!d{\bf k}\,
  b_{\bf k}^*b_{\bf -k}^*\e^{2i\omega_kT_f}+
  \int\!d{\bf k}\,\sqrt{2\omega_{\bf k}}b_{\bf k}^*\phi_f({\bf k})
  \e^{i\omega_kT_f}-
  {1\over2}\int\!d{\bf k}\,\omega_{\bf k}\phi_f({\bf k})\phi_f({\bf -k})
\]
where $T_f$ is some final time moment (the limit $T_f\to+\infty$ is
assumed), while $\phi_i({\bf k})$ and $\phi_f({\bf k})$ are the
Fourier transformations of the field in the initial and final
asymptotic regions.

It is easy to see that if one takes $j$ to be of order
$1/\sqrt{\lambda}$, then the integral (\ref{int_repr}) has the
semiclassical nature in the limit $\lambda\to0$, $\lambda E$, $\lambda
n=$ fixed, and one can expect that it is saturated by a saddle point,
where $\phi$, $\phi'$, $b_{\bf k}$ and $b^*_{\bf k}$ are of order
$1/\sqrt{\lambda}$, and $\xi$ and $\eta$ are of order 1.  In this
case, the cross section certainly has the exponential form
(\ref{intro_exp}).  However, the assumption $j\sim1/\sqrt{\lambda}$
contradicts the requirement that $j$ does not depend on $\lambda$,
which forces us to take $j$ parametrically smaller, $j\sim1$.  This
difficulty is obviously the consequence of the non--semiclassical
nature of the initial state that contains few energetic particles, in
contrast with the final state consisting of many soft ones.  A method
to overcome this difficulty has been suggested for the instanton
transitions \cite{RT,T} and can be applied for our problem in an
analogous way.  In this method one evaluates the integral
(\ref{int_repr}) for $j=\alpha/\sqrt{\lambda}$, where $\alpha$ is some
constant, in saddle--point approximation, and find the cross section
in the exponential form $\sigma\sim\e^W$, where $W$ is of order
$1/\lambda$ and depends on $\alpha$.  After that one takes the limit
$\alpha\to0$.  The claim is that in this limit one reproduces the
value of $W$ at $j\sim1$.

Apparently, the weak point in this way of reasoning is the limit
$\alpha\to0$: it is neither obvious that this limit is smooth, nor
that it reproduces the amplitude with one initial particle.  Moreover,
the exponent of the cross section of the $1\to n$ process contains
contribution from {\em loops} (the term $2\mbox{Re}B\lambda n^2$ in
eq.(\ref{Fpert})) that seems to be of quantum, rather than classical,
nature.  Nevertheless, it turns out that this nontrivial exponentiated
loops are reproduced by the semiclassical calculations (see Sect.4
below), which is a sound argument in favor to the hypothesis about the
$\alpha\to0$ limit.

Therefore, we assume that $j\sim1/\sqrt{\lambda}$ and look for the
saddle point of the exponent of the integrand in (\ref{int_repr}).
The saddle--point equations can be divided into two groups.  The first
group contains the equations for $\phi$,
\begin{equation}
  {\delta S\over\delta\phi}=ij\delta(x)
  \label{phi_fe}
\end{equation}
\begin{equation}
  i\dot{\phi}_i({\bf k})+\omega_{\bf k}\phi_i({\bf k})=0
  \label{phi_in}
\end{equation}
\begin{equation}
  i\dot{\phi}_f({\bf k})-\omega_{\bf k}\phi_f({\bf k})+
  \sqrt{2\omega_{\bf k}}b^*_{\bf -k}\e^{i\omega_kT_f}=0
  \label{phi_fin1}
\end{equation}
\begin{equation}
  -b_{\bf k}\e^{i\omega_ k\xi+i\eta}-
  b_{\bf -k}^*\e^{2i\omega_kT_f}+
  \sqrt{2\omega_{\bf k}}\phi_f({\bf k})\e^{i\omega_kT_f}=0
  \label{phi_fin2}
\end{equation}
and similar equations for $\phi'$.  The equations from the second
group relate the energy $E$ and the number of final particles $n$ to
other parameters,
\begin{equation}
  E=\int\!d{\bf k}\,\omega_{\bf k}b^*_{\bf k}b_{\bf k}
    \e^{i\omega_k\xi+i\eta}
  \label{Ebb*}
\end{equation}
\begin{equation}
  n=\int\!d{\bf k}\,b^*_{\bf k}b_{\bf k}\e^{i\omega_k\xi+i\eta}
  \label{nbb*}
\end{equation}

First, let us consider the equations for $\phi$.  Eq.(\ref{phi_fe}) is
simply the field equation with a $\delta$--like source, while
eqs.(\ref{phi_in}), (\ref{phi_fin1}), (\ref{phi_fin2}) can be
considered as boundary conditions in initial and final asymptotics
$t\to\mp\infty$.  It is convenient to rewrite these boundary
conditions in a more transparent form.  For this end we note that
since $\phi$ is a superposition of plane waves in the limit
$t\to-\infty$ and $t\to+\infty$, eq.(\ref{phi_in}) can be satisfied
only if the initial asymptotics of $\phi$ is purely Feynman,
\begin{equation}
  \phi_i({\bf k})=a^*_{\bf -k}\e^{i\omega_kt},\qquad t\to-\infty
  \label{phi_in_as}
\end{equation}
where $a_{\bf k}$ are arbitrary Fourier components, while
eq.(\ref{phi_fin2}) implies the following asymptotics of $\phi$ in the
final asymptotic region,
\begin{equation}
  \phi_f({\bf k})={1\over\sqrt{2\omega_{\bf k}}}\left(
  b_{\bf k}\e^{i\omega_k\xi+i\eta-i\omega_kt}+
  b^*_{\bf -k}\e^{i\omega_kt}\right),\qquad t\to+\infty
  \label{phi_fin_as}
\end{equation}
It is easy to see that eq.(\ref{phi_fin_as}) satisfies the condition
(\ref{phi_fin1}) automatically.

Let us turn to the equations from the second group.  The physical
meaning of eqs.(\ref{Ebb*}) and (\ref{nbb*}) is simple: they read that
$E$ and $n$ are the energy and the number of particles of the field
$\phi$ in its final asymptotics (\ref{phi_fin_as}).  Since $\phi$
satisfies the sourceless field equation at all values of $t$ but $t=0$
(where the source is not vanishing), the energy of the field conserves
in the two regions $t>0$ and $t<0$ separately, but may have
discontinuity at $t=0$.  The energy in the region $t>0$ is equal to
that in the limit $t\to+\infty$ and therefore is $E$.  To find the
energy at $t<0$ we make use of the $\phi$'s initial asymptotics, and
since the latter contains only Feynman components, it vanishes.  So,
we find that the energy has a finite jump at $t=0$ which, naturally,
is associated with the $\delta$--functional source located at $t={\bf
x}=0$.

To simplify further discussions, let us make two conjectures that, as
we will see in what follows, do not lead to contradiction.  The first
is that the saddle--point values of $b_{\bf k}$ and $b^*_{\bf k}$ are
complex conjugated to each other.  The physical meaning of this
assumption is that the sum over final states in eq.(\ref{sigma}) is
saturated by a single coherent state.  The second conjecture is that
the saddle point values of $\xi$ and $\eta$ are purely imaginary,
\[
  \xi=-iT,\qquad \eta=i\theta
\]
where $T$ and $\theta$ are real (for further convenience we choose
different sign conventions for $T$ and $\theta$).

With the two conjectures formulated above, the field configuration
describing the multiparticle process at given energy $E$ and
multiplicity $n$ is the solution to the field equation with source
\begin{equation}
  {\delta S\over\delta\phi}=ij\delta^{d+1}(x)
  \label{14*}
\end{equation}
with the two boundary conditions,
\begin{equation}
  \phi({\bf k})=a_{\bf k}\e^{i\omega_kt},\qquad t\to-\infty
  \label{14**}
\end{equation}
\begin{equation}
  \phi({\bf k})={1\over\sqrt{2\omega_{\bf k}}}\left(
  b_{\bf k}\e^{\omega_kT-\theta-i\omega_kt}+
  b_{\bf -k}^*\e^{i\omega_kt}\right),\qquad t\to+\infty
  \label{phi_bc}
\end{equation}
It is easy to notice that the initial boundary conditions (\ref{14**})
can be reformulated in Euclidean language.  In fact, making the Wick
rotation to the Euclidean time $\tau=-it$, eq.(\ref{14**}) reads that
$\phi({\bf k})$, as a function of $\tau$, contains only the decaying
component in the asymptotics $\tau\to+\infty$,
\begin{equation}
  \phi({\bf k})=a_{\bf k}\e^{-\omega_k\tau},\qquad\tau\to+\infty
  \label{in_as}
\end{equation}
In contrast, the the final asymptotics (\ref{phi_bc}) contains both
frequencies and cannot be rewritten in Euclidean language.  Moreover
the field in the final asymptotics is, in general, complex.

The boundary value problem for $\phi'$ can be derived in analogous
way,
\[
  {\delta S\over\delta\phi}=-ij\delta^{d+1}(x)
\]
\[
  \phi'({\bf k})=a_{\bf k}\e^{-i\omega_kt},\qquad t\to-\infty
\]
\begin{equation}
  \phi'({\bf k})={1\over\sqrt{2\omega_{\bf k}}}\left(
  b_{\bf k}\e^{-i\omega_kt}+
  b_{\bf -k}^*\e^{\omega_kT-\theta+i\omega_kt}\right),
  \qquad t\to+\infty
  \label{phi'_bc}
\end{equation}
Notice that if $\phi$ is a solution to the boundary value problem
(\ref{14*}, \ref{14**} \ref{phi_bc}), its complex conjugate $\phi^*$
satisfies eqs.(\ref{phi'_bc}).  This fact simplifies our calculations,
since we need to solve only one boundary value problem instead of two.
In further discussions we will assume that $\phi'=\phi^*$.

The saddle point of the integral (\ref{int_repr}) determines the cross
section, which has the exponential form,
\[
  \sigma(E,n)\sim\e^{W(E,n)}
\]
where
\begin{equation}
  W(E,n)={1\over\lambda}F(\lambda n,\epsilon)
  =ET-n\theta-2\mbox{Im}S[\phi]
  \label{W}
\end{equation}

The relation between $E$, $n$ and $T$, $\theta$, can be found either
from eqs.(\ref{Ebb*}) and (\ref{nbb*}), or, equivalently, from
\begin{equation}
  2{\partial\mbox{Im}S\over\partial T}=E,\qquad
  2{\partial\mbox{Im}S\over\partial\theta}=n.
  \label{dS/dT}
\end{equation}
which can be easily understood if one recalls that $\xi=-iT$ and
$\eta=i\theta$ are the saddle point of the integrand in
eq.(\ref{int_repr}).  From (\ref{W}) one sees that $W(E,n)$ is the
Legendre transformation of $2\mbox{Im}S(T,\theta)$, and therefore one
obtains the following important relations,
\begin{equation}
  {\partial W\over\partial E} = T,\qquad
  {\partial W\over\partial n} = -\theta
  \label{dW/dn}
\end{equation}

Having derived the boundary value problem for calculating the
transition rate at finite $j$, let us discuss the limit $j\to0$.  It
can be shown that in this limit the field configuration becomes
singular at $x=0$.  In fact, according to eq.(\ref{14*}), $\phi$ has
discontinuity at $t=0$,
\[
  \delta\dot{\phi}({\bf x})=
  \dot{\phi}({\bf x})|_{t=+0}-\dot{\phi}({\bf x})|_{t=-0}=
  ij\delta^d({\bf x})
\]
This discontinuity leads a jump of the energy at $t=0$, since the
latter contains the term ${1\over2}\int\!d{\bf x}\,\dot{\phi}^2$.  On
the other hand, the discontinuity of the energy $E$ is supposed to be
finite while that of $\dot{\phi}$ is proportional to $j$ and tends to
0.  When $j$ is small one has
\[
  E=\int\!d{\bf x}\,\dot{\phi}(0,{\bf x})\delta\dot{\phi}(0,{\bf x})
   =ij\dot{\phi}(0)
\]
One sees that when $E$ is fixed and $j\to0$, $\dot{\phi}(0)$ goes to
infinity, which means that the field configuration becomes singular at
$x=0$ in the limit of vanishing source.  This is not surprising, since
in the limit $j\to0$ eq.(\ref{phi_fe}) becomes the sourceless field
equation, whose regular solutions conserve the energy and therefore do
not obey the boundary conditions.

So, to evaluate the transition rate one should find the solution to
the field equation which obeys the boundary conditions and has
singularity at $t={\bf x}=0$, but remains regular elsewhere in the
Minkowskian space--time.  For doing calculations, however, we will use
another formulation of the boundary value problem.

\subsection{Extremization procedure}

Let us discuss the structure of singularities of our solution.  Recall
that in Minkowskian space--time, $\phi$ is regular everywhere except a
point--like singularity at $x=0$.  However, if one extrapolates $\phi$
into Euclidean times, it may occur that $\phi$ develops more
singularities beside that at $x=0$.  Let us consider a simple
possibility that $\phi$ is singular on some $d$--dimensional surface
$A$ in the Euclidean space--time, which we will parametrize either as
$x_{\mu}=x_{\mu}(s_i)$, where $s_i$ are $d$ coordinates on the
surface, or $\tau=\tau_0({\bf x})$.  In the region near $A$, $\phi$ is
inverse to the distance to the surface,
\begin{equation}
  \phi\sim\sqrt{2\over\lambda}{1\over l(x)}
  \label{leadsing}
\end{equation}
where $l(x)$ is the distance form the point $x$ to $A$.  In fact
eq.(\ref{leadsing}) is the only possible form of the leading
singularity of $\phi$ in the region near the singularity surface $A$.
In fact, one can even develop a perturbation theory on $l$ and find
the correction to eq.(\ref{leadsing}), which is of order $l(x)$.
However, one will see soon that ambiguity begins at the order of
$O(l^3)$ (on other words, the terms higher than $O(l^3)$ is not
defined uniquely by the form of $A$), which reflect the fact that the
solution is not defined uniquely by the surface where it is singular.
If two solutions are singular on the same surface $A$, the difference
between them goes to 0 when one approaches the surface as $l^3$.  Note
that if $A$ touches the plane $\tau=0$ only at one point $x=0$, then
in Minkowskian space--time $\phi$ has the required structure, i.e. is
singular only at $x=0$.

Let us take an arbitrary surface $A$ satisfying the latter requirement
and determine a field configuration $\phi$, which consists of two
parts $\phi_1$ and $\phi_2$, as follows.  Both $\phi_1$ and $\phi_2$
are supposed to be solutions to the field equation and singular on
$A$, but each of them obeys one boundary condition from
eqs.(\ref{14**}) and (\ref{phi_bc}).  Namely, we require that $\phi_1$
decreases in the Euclidean asymptotics,
\[
  \phi_1\to0,\qquad\tau\to+\infty
\]
while $\phi_2$ obeys the boundary condition in the Minkowskian limit,
\[
  \phi_2({\bf k})={1\over\sqrt{2\omega_{\bf k}}}\left(
  b_{\bf k}\e^{\omega_kT-\theta-i\omega_kt}+
  b_{\bf -k}^*\e^{i\omega_kt}\right),\qquad t\to+\infty
\]
One may imagine the field $\phi$ is defined on a particular contour on
the complex time plane, which at each value of ${\bf x}$ goes along
the Euclidean time axis from $i\infty$ to some $i\tau_0({\bf x})$
lying on the singularity surface and then goes back to 0 and then
along the Minkowskian time axis to $\infty$ (fig.\ref{contour}).  The
field $\phi_1$ is defined on the first part of the contour,
$(i\infty,i\tau_0)$, while $\phi_2$ on the two final parts,
$(i\tau_0,0)$ and $(0,\infty)$.  Note that there is a region on the
Euclidean time axis where both $\phi_1$ and $\phi_2$ are defined,
namely $(i\tau_0,0)$.

Despite the fact that $\phi$ obeys the boundary conditions of the
boundary value problem, it is not the solution to the latter, since
$\phi_1$ and $\phi_2$ need not to be equal at $t=0$.  In other words,
for a generic surface $A$, $\phi$ contains discontinuities on the
whole plane $t=0$, instead of being singular at one point $t={\bf
x}=0$.  On the other hand, if one manages to choose the surface $A$ in
such a way that $\phi_1(0,{\bf x})=\phi_2(0,{\bf x})$ for any ${\bf
x}\neq0$, $\phi$ would be the solution to the boundary value problem.

Let us define the Euclidean action of $\phi$ as the sum of the action
of $\phi_1$ and $\phi_2$, $S[\phi_1]$ and $S[\phi_2]$, each calculated
along the corresponding part of the contour,
\[
  S_E[\phi]=-\int\limits_{+\infty}^{\tau_0({\bf x})}\!d\tau d{\bf x}
  \left({1\over2}(\partial_\mu\phi_1)^2+V(\phi_1)\right)-
  \int\limits_{\tau_0({\bf x})}^0\!d\tau d{\bf x}
  \left({1\over2}(\partial_\mu\phi_2)^2+V(\phi_2)\right)-
  i\int\limits_0^{\infty}\!dt d{\bf x}\,L(\phi_2)
\]
Since $\phi_{1,2}$ are singular at $\tau=\tau_0({\bf x})$, both
$S[\phi_1]$ and $S[\phi_2]$ are infinite, but since the integration
contour for $\phi_1$ and $\phi_2$ goes at different directions near
the singularities, one can hope that their sum $S[\phi]$ is
nevertheless finite.

Now we will show that the ``correct'' singularity surface $A$
determined by the condition that $\phi_1=\phi_2$ at $t=0$ can be found
by extremizing the real part of the Euclidean action $S_E[\phi]$ with
respect to all possible form of the surface $A$, with the requirement
that the point $\tau={\bf x}=0$ lies on the latter.  First let us
regularize $\phi$ to avoid dealing with infinities in intermediate
calculations.  For this end we replace the condition that $\phi$ is
singular on $A$ by the condition that $\phi=\phi_0$ on the same
surface, where $\phi_0$ is some large, but finite number, which will
eventually tends to infinity.  So, we set $\phi_1=\phi_2=\phi_0$ on
$A$.  Since $\phi_1$ and $\phi_2$ are, in general, different in the
region near the surface, we expect that the derivatives of
$\phi_{1,2}$ are different on $A$.  Let us denote
\begin{equation}
  \partial_n(\phi_1-\phi_2)=j(s_i)
  \label{page11}
\end{equation}
where $\partial_n$ is the derivative along the direction normal to
$A$.  The configuration $\phi$, thus, can be regarded as the solution
to the field equation with a source that is distributed over the
surface $A$,
\[
  {\partial S_E\over\partial\phi}=j(x)=\int\!ds_i\,
  j(s_i)\delta(x-x_\mu(s))
\]
(in this formula we assume that the appropriate metric factor has been
included in $ds_i$).  Now, as an intermediate step, we show that once
$S_E[\phi]$ has been extremized with respect to $A$, the source $j(x)$
is proportional to $\delta(x)$ (in other words, the source is located
at the point $x=0$ rather than distributed over $A$).  Let us take an
arbitrary singularity surface $A$ and deform it slightly to $A'$.
This can be represented as shifting each point $x_\mu$ on $A$ by a
small vector $\delta x_\mu=n_\mu\delta x$, where $n_\mu$ is the unit
vector perpendicular to $A$, so that $x_\mu+\delta n_\mu$ lies on
$A'$.  To ensure that $x=0$ is always a singular point, we require
that $\delta x_\mu|_{x=0}=0$ (fig.\ref{shift}).

The new surface $A'$ thus corresponds to new configurations
$\phi_{1,2}'=\phi_{1,2}+\delta\phi_{1,2}$.  We will evaluate the
variation of $S[\phi_1]$ and $S[\phi_2]$ separately.  The variation of
$S[\phi_1]$ is due to two factors: the first is the variation of the
field, $\phi_1\to\phi_1+\delta\phi_1$ and the second is the change of
the integration region.  The first contribution can be reduced to a
boundary integral, since $\phi_1$ is a solution to the field equation,
\[
  -\!\!\int\limits_{\infty}^{\tau_0({\bf x})}\!\!\!d\tau d{\bf x}
  \delta\left({1\over2}(\partial_\mu\phi_1)^2+V(\phi_1)\right)
  =-\int\limits_A\!dx\,
  [\partial_\mu\phi_1\partial_\mu\delta\phi_1-V'(\phi_1)\delta\phi_1]=
  -\int\limits_{A}\!ds\,(\partial_\mu\phi_1\cdot n_\mu)\delta\phi_1
\]
If $\delta x_\mu$ is small, the second contribution that is associated
with the change of the integration region can be also reduced to a
boundary integral, where each point on the surface $A$ is integrated
with the weight $\delta x$.  Therefore, the full variation of
$S[\phi_1]$ is
\[
  \delta S_E[\phi_1]=\int\limits_A\!ds\,
  \left[(\partial_\mu\phi\cdot n_\mu)\delta\phi
  -\left({1\over2}(\partial_\mu\phi_1)^2+V(\phi_1)\right)\delta x(s)\right]
\]
Let us make use of the fact that $\phi_1=\phi_0$ on $A$, and
$\phi_1'=\phi_0$ on $A'$.  We have
\[
  \delta\phi(x_\mu)=\phi'(x_\mu)-\phi(x_\mu)=
  \phi'(x_\mu)-\phi'(x_\mu+\delta x_\mu)=
  -(\partial_\mu\phi\cdot n_\mu)\delta x(s)
\]
Therefore,
\begin{equation}
  \delta S[\phi_1]=\int\limits_A\!ds\,\left[(\partial_\mu\phi_1)^2-
  \left({1\over2}(\partial_\mu\phi_1)^2+V(\phi_1)\right)
  \right]\delta x(s)
  \label{deltaS1}
\end{equation}
\[
  =\int\limits_A\!ds\,\left[{1\over2}(\partial_n\phi_1)^2-V(\phi_1)\right]
\]
where we have made use of the fact that on $A$ the derivatives of
$\phi$ along directions tangent to $A$ vanish (since $\phi$ is
constant on $A$).

Let us turn to the variation of $S[\phi_2]$.  The computation is
completely analogous to the case of $S[\phi_1]$, with the exception
that now there is a boundary term at $t=+\infty$.  So we have,
\begin{equation}
  \delta S_E[\phi_2]=-\int\!ds\,
  \left[{1\over2}(\partial_n\phi_2)^2-V(\phi_2)\right]
  -i\int\!d{\bf x}\,\partial_0\phi_2\delta\phi_2|_{t=+\infty}
  \label{deltaS2}
\end{equation}
However one can show that the boundary term at $t=+\infty$ is purely
imaginary.  In fact, since $\phi_2$ and $\phi_2'$ obey the boundary
condition (\ref{phi_bc}) with the same $T$ and $\theta$, we have at
$t\to\infty$,
\[
  \delta\phi_2({\bf k})={1\over\sqrt{2\omega_{\bf k}}}\left(
  \delta b_{\bf k}\e^{\omega_kT-\theta-i\omega_kt}+
  \delta b_{\bf -k}^*\e^{i\omega_kt}\right),\qquad t\to+\infty
\]
Substituting this, as well as the asymptotics for $\phi_2$, to the
boundary term at $t=+\infty$, we see that the latter term in the r.h.s.
of eq.(\ref{deltaS2}) is in fact purely imaginary
\[
  i\int\!d{\bf x}\,\partial_0\phi_2\delta\phi_2=
  \int\!{d{\bf k}\over(2\pi)^d}\,
  \left(b_{\bf k}\delta b_{\bf k}^*-b^*_{\bf k}\delta b_{\bf k}\right)
  \e^{\omega_kT-\theta}
\]
and since we are interested only in the real part of the Euclidean
action, this term can be dropped it in further calculations.  Taking the
sum of (\ref{deltaS1}) and (\ref{deltaS2}), one finds,
\[
  \delta\mbox{Re} S_E[\phi]={1\over2}\int\limits_A\!ds\,\left(
  (\partial_n\phi_1)^2-(\partial_n\phi_2)^2\right)\delta x(s)
\]
Now when one takes $\phi_0$ to be large, both $\partial_n\phi_{1,2}$
are large but the difference between them is small.  Making use of
eq.(\ref{page11}) one writes
\[
  \delta\mbox{Re} S[\phi]=\int\!ds\,(\partial_n\phi)\delta x(s)j(s)
\]
Now we see that the requirement that $\delta\mbox{Re}S[\phi]=0$ for
all variations $A$ obeying $\delta x|_{x=0}=0$ can be satisfied only
if $j(s)$ is proportional to the delta function, $j(x)=j_0\delta(x)$.
So, when we extremize the real part of the action $S_E[\phi]$, varying
the surface $A$, the source $j$, which is at first distributed over
$A$, gather to a localized delta--functional source
$j(x)=j_0\delta(x)$

Suppose that we have performed this extremization procedure.  Since
now $\phi_1$ and $\phi_2$ are equal on $A$ and there normal
derivatives are also equal (except from the point $x=0$), these fields
coincide in the region where they are both determined, namely, at
$\tau_0({\bf x})<\tau<0$.  In particular, $\phi_1=\phi_2$ everywhere
on the plane $\tau=0$ but $\tau={\bf x}=0$.

So far we have been dealing with the regularized field configurations.
Let us now take the limit $\phi_0\to\infty$.  The coincidence of
$\phi_1$ and $\phi_2$ at $t=0$ remains in this limit, however now the
point $t={\bf x}=0$ becomes singular.  The strength of the
delta--functional source, $j_0$, goes to 0 in order to keep the jump
of energy finite.  So, $\phi$ obeys the sourceless field equation in
Minkowskian space--time, and is singular only at $x=0$, thus it is the
solution to the boundary value problem.

To summarize, we have shown that the solution to the boundary value
problem can be found by extremizing the real part of the Euclidean
action $S_E[\phi]$ (or the imaginary part of the Minkowskian action)
over all singularity surfaces $A$ containing the point $t={\bf x}=0$.
Let us now apply this formalism to find the cross section in various
limiting cases.

\section{Tree--level cross sections}

\subsection{General consideration}

Consider the exponent for the cross section, $F(\lambda n,\epsilon)$,
at small $\lambda n$.  Keeping in mind the formula (\ref{Fpert}), one
expect that at $\epsilon\sim1$, $\lambda n\ll1$, the function $F$ has
the following form,
\begin{equation}
  F(\lambda n,\epsilon)=\lambda n\ln{\lambda n}-\lambda n+
  \lambda nf(\epsilon)+O(\lambda^2n^2)
  \label{Ftree}
\end{equation}
where $f(\epsilon)$ is some function of $\epsilon$.  In what follows
we will neglect the terms $O(\lambda^2n^2)$ and higher.  Since these
terms come from loops, it is equivalent to considering the tree level.
Another way to see this is to compute the cross section corresponding
to (\ref{Ftree}),
\begin{equation}
  \sigma(E,n)\sim\exp\left({1\over\lambda}F(n\lambda,\epsilon)\right)
  \sim n!\lambda^n\e^{nf(\epsilon)}
  \label{simple}
\end{equation}
We see that the cross section depends on the coupling constant
$\lambda$ as $\lambda^n$, which is natural for tree diagrams whose
number of vertices is $n/2$.  Even at tree level, the cross section at
arbitrary $\epsilon$ has not been calculated (for lower bound see
ref.\ \cite{Vol_bound}).  In the non--relativistic limit
$\epsilon\ll1$ the perturbative result (\ref{Fpert}) yields the
following formula for $f(\epsilon)$,
\begin{equation}
  f(\epsilon)=-\ln16+{d\over2}\ln{\epsilon\over\pi d}+{d\over2}
  +\left({d-2\over4}-{5\over3}\right)\epsilon+O(\epsilon^2)
  \label{fepsilon}
\end{equation}
In this section we make no attempt to compute $f(\epsilon)$ at
arbitrary value of $\epsilon$.  Our main goal is to reproduce
eq.(\ref{fepsilon}) from the formalism of Sect.2.  We will also try to
estimate the tree cross section from below in the limit
$\epsilon\to\infty$.

Before considering the small--$\epsilon$ limit, let us first point out
that the calculation procedure can be considerably simplified if one
restricts himself to the tree level.  We will work in the
``extremization'' formalism, not in the framework of the original
Minkowskian boundary value problem, so we take an arbitrary surface
$A$ and calculate $S_E[\phi]$ (we will deal only with the Euclidean
action, so for simplicity further we will drop the index $E$).

First note that from eqs.(\ref{dW/dn}) and (\ref{Ftree}) one finds
\[
  \theta=-{1\over\lambda}{\partial F\over\partial n}
  =-\ln{\lambda n\over16}-f(\epsilon)
\]
So, in the limit $\lambda n\to0$, $\theta\gg1$ independent of the form
of the function $f(\epsilon)$.  We see that in the final asymptotics
the Feynman part of $\phi$, $b_{\bf
k}\e^{-i\omega_kt+\omega_kT-\theta}$ is much smaller than the
anti--Feynman part, $b^*_{\bf k}\e^{i\omega_kt}$.

Recall that the initial asymptotics of $\phi_1$ is the same as that of
$\phi$ eq.(\ref{in_as}).  By construction, $\phi_1$ is defined on the
first part of the contour of fig.\ref{contour}, $(i\infty,0)$, however
one can always analytically continue $\phi_1$ to other parts.  Let us
investigate the difference between $\phi_1$ and $\phi_2$ on the parts
2 and 3.  Denote
\[
  \tilde{\phi}=\phi_2-\phi_1
\]
so $\tilde{\phi}$ is regular (in fact, equal to 0) on the surface $A$
and obeys the boundary condition
\[
  \tilde{\phi}({\bf k})={1\over\sqrt{2\omega_{\bf k}}}\left(
  b_{\bf k}\e^{-i\omega_kt+\omega_kT-\theta}+
  (b^*_{\bf -k}-a^*_{\bf -k})\e^{i\omega_kt}\right)
\]
in the final asymptotics.  Let us show that the anti--Feynman part of
$\tilde{\phi}$, $b^*_{\bf -k}-a^*_{\bf -k}$ is small and proportional
to $\e^{-\theta}$.  Supposing that this is true, then the whole
$\tilde{\phi}$ is a small perturbation on $\phi_1$ and thus obeys the
linearized equation
\begin{equation}
  (\partial_{\mu}^2+1+3\lambda\phi_1^2)\tilde{\phi}=0
  \label{lin_eq}
\end{equation}
with two boundary conditions
\[
  \tilde{\phi}|_{A}=0
\]
\begin{equation}
  \tilde{\phi}(k)={1\over\sqrt{2\omega_k}}\,a_{\bf k}
  \e^{-i\omega_kt+\omega_kT-\theta}+~
  \mbox{any anti--Feynman part},\qquad t\to+\infty
  \label{asympt2}
\end{equation}
(we have made use of the assumption that $b_{\bf k}\approx a_{\bf
k}$).  Since the equation is linear and the final boundary conditions
contains a factor of $\e^{-\theta}$, the solution $\tilde{\phi}$ is
also proportional to $\e^{-\theta}$, which is consistent with our
starting assumption.

Now making use of the smallness of $\tilde{\phi}$, the action, up to
the contributions of order $\e^{-\theta}$, is equal to
\begin{equation}
  S[\phi]=S_1[\phi_1]+S_{\{2,3\}}[\phi_1]-\int_{\{3\}}\!dx\,\left[
  (\partial_\mu\phi_1)(\partial_\mu\tilde{\phi})-
  V'(\phi_1)\tilde{\phi}\right]
  \label{S123}
\end{equation}
It is easy to see that the action of $\phi_1$ on the whole contour
(the first two term in the r.h.s. of eq.(\ref{S123})) vanishes (one
can see this, for example, by making the Wick rotation of the third
part of the contour).  The last term in eq.(\ref{S123}) is reduces to
boundary integrals.  The boundary term on $A$ is equal to 0 since near
$A$ $\tilde{\phi}(x)$ tends to 0 as $l^3$, where $l$ is the distance
from $x$ to $A$, while $\partial\phi_1\sim l^{-2}$.  The boundary term
at $t=+\infty$ is
\begin{equation}
  S[\phi]=i\int\!d{\bf x}\,\tilde{\phi}\partial_t\phi_1|_{t=+\infty}
  ={1\over2}\int\!d{\bf k}\,a^*_{\bf k}\,
  a_{\bf k}\e^{\omega_kT-\theta}
  \label{Stree}
\end{equation}
So, to compute the action, one need not really solve
eq.(\ref{lin_eq}): only the knowledge of $\phi_1$ (more precisely, its
Euclidean asymptotics) is required.  Note that the Euclidean action is
real.

Once the action is found, one should extremize (\ref{Stree}) over all
solutions that are singular at $x=0$.  In fact, one can see that one
should extremize only $\int\!d{\bf k}\,a^*_{\bf k}a_{\bf
k}\e^{\omega_kT}$, since $\e^{-\theta}$ is just an overall factor in
eq.(\ref{Stree}).  Moreover, since the action is bounded from below by
0, the extremum of the action is most likely the true minimum.  We
also expect that the action contains a classical factor $1/\lambda$,
so, the extremized action has the following form
\begin{equation}
  2S(T,\theta)={1\over\lambda}\e^{g(T)-\theta}
  \label{g_def}
\end{equation}
where $g(T)$ is some function.  Eqs.(\ref{dS/dT}) then read
\begin{equation}
  E=2{\partial S\over\partial T}={1\over\lambda}g'(T)\e^{g(T)-\theta}
  \label{Etree}
\end{equation}
\begin{equation}
  n=-2{\partial S\over\partial\theta}={1\over\lambda}\e^{g(T)-\theta}
  \label{ntree}
\end{equation}
These equations should be solved with respect to $T$ and $\theta$.
Dividing (\ref{Etree}) to (\ref{ntree}), one obtains
\begin{equation}
  1+\epsilon=g'(T)
  \label{Teps}
\end{equation}
We see that the parameter $T$ depends on $\epsilon$ but not on
$\theta$.  Regarding (\ref{Teps}) as an equation on $T$, we denote its
solution as
\[
  T=T(\epsilon)
\]
and from eq.(\ref{ntree}) one finds $\theta$ as a function of
$\epsilon$ and $n$,
\[
  \theta=g(T(\epsilon))-\ln(\lambda n)
\]
Let us substitutes the solution to the exponent of the cross section.
One obtains
\[
  {1\over\lambda}W=ET-n\theta-2S(T,\theta)=
  n(1+\epsilon)T(\epsilon)-n(g(T(\epsilon))-\ln\lambda n)-n
\]
Comparing the last equation with (\ref{Ftree}), one finds
\begin{equation}
  f(\epsilon)=(1+\epsilon)T(\epsilon)-g(T(\epsilon))
  \label{fefin}
\end{equation}

Therefore, the problem of finding the tree cross section at any value
of $E$ and $n$ can be formulated entirely in the Euclidean
space--time.  One looks for all solutions $\phi_1(\tau,{\bf x})$ to
the Euclidean field equations which are singular at $\tau={\bf x}=0$
and decay as $\tau\to+\infty$, and calculates for each solution the
corresponding Fourier components $a_{\bf k}$ from its asymptotics at
$\tau\to\infty$.  Then one should maximize the integral $\int\!d{\bf
k}\,a^*_{\bf k}a_{\bf k}\e^{\omega_kT}$ and determines the function
$g(T)$ (eq.(\ref{g_def})).  The required $f(\epsilon)$ can be
calculated using eq.(\ref{fefin}), where $T(\epsilon)$ is the solution
to eq.(\ref{Teps})

Unfortunately, we are unable to carry out this program for arbitrary
values of $\epsilon$ due to the non--linearity of the field equation.
At small $\epsilon$ (which corresponds to non--relativistic final
states) we will see that it reproduces the result found by
perturbative calculations.

\subsection{Non--relativistic regime}

\subsubsection{Leading order}

At small $\epsilon$, the typical momentum of final particles is
$k_0=\epsilon^{1/2}$ of much smaller than the mass, so one can expect
that $\phi_1$ is a slowly varying function of ${\bf x}$.  The ${\bf
x}$--independent solution to the field equation which is singular at
$\tau=0$ and decays as $\tau\to\infty$ is known explicitly,
\begin{equation}
  \phi_1(\tau)=\sqrt{2\over\lambda}\,{1\over\sinh\tau}\, ,
  \label{homo}
\end{equation}
Let us restrict ourselves to the leading order.  The field
configuration in the first approximation can be obtained from
(\ref{homo}) by a simple modification,
\begin{equation}
  \phi_1(\tau,{\bf x})=\sqrt{2\over\lambda}\,
  {1\over\sinh(\tau-\tau_0({\bf x}))}
  \label{leading}
\end{equation}
where $\tau_0({\bf x})$ is a slowly varying function of ${\bf x}$.
One can check that eq.(\ref{leading}) satisfies the field equation to
the accuracy of $O((\partial_{\bf x}\tau_0)^2)$.  Note that
$\tau=\tau_0({\bf x})$ is the surface of singularities of $\phi_1$.
The point $\tau={\bf x}=0$ should lie on the latter, so we require
that $\tau_0(0)=0$.

First, from eq.(\ref{leading}) one can relate $a_{\bf k}$ to
$\tau_0(x)$.  At $\tau\to\infty$, the asymptotics of $\phi_1$ in
eq.(\ref{leading}) is
\begin{equation}
  \phi(\tau,{\bf x})=\sqrt{8\over\lambda}\,
  \e^{\tau_0({\bf x})}\e^{-\tau}
  \label{equal}
\end{equation}
On the other hand, $\phi_1$ can be expanded into plane wave in this
region,
\begin{equation}
  \phi_1=\int\!{d{\bf k}\over(2\pi)^{d/2}\sqrt{2\omega_k}}\,a_{\bf
  -k} \e^{i{\bf kx}-\omega_k\tau}
  \label{equal'}
\end{equation}
Recalling that typical momentum ${\bf k}$ is small, as the first
approximation one can replace $\omega_{\bf k}$ by 1 and the r.h.s. of
eq.(\ref{equal'}) behaves like $\e^{-\tau}$, as that of
eq.(\ref{equal}).  Comparing eqs.(\ref{equal}) and (\ref{equal'}), one
obtains,
\begin{equation}
  \int\!{d{\bf k}\over(2\pi)^{d/2}}\,a_{\bf k}\e^{i{\bf kx}}=
  \sqrt{16\over\lambda}\,\e^{\tau_0({\bf x})}
  \label{btau}
\end{equation}
Taking in this equation ${\bf x}=0$, and recalling the requirement
$\tau({\bf x})=0$, one finds the following constraint on $a_{\bf k}$,
\begin{equation}
  \int\!{d{\bf k}\over(2\pi)^{d/2}}\,a_{\bf k}=
  \sqrt{16\over\lambda}
  \label{constraint}
\end{equation}

To find $W$ one should extremize the r.h.s of eq.(\ref{Stree}) with
the constraint (\ref{constraint}).  We follow the standard
technique and introduce the term
\[
  C\left(\int\!{d^d{\bf k}\over(2\pi)^{d/2}}\,a_{\bf k}-(2\pi)^{d/2}
  \sqrt{16\over\lambda}\right)
\]
where $C$ is the Lagrange multiplier to the r.h.s. of eq.(\ref{Stree})
and, varying with respect to $a^*_{\bf k}$, we obtain $a_{\bf k}$,
\begin{equation}
  a_{\bf k}=C\e^{-\omega_kT}
  \label{a}
\end{equation}
It is easy to notice that (\ref{a}) in fact the minimizes of the
action.  The constant $C$ can be determined from
eq.(\ref{constraint}),
\[
  C=\sqrt{16\over\lambda}T^{d/2}e^T
\]
Let us now calculate the function $g(T)$.  According to
eq.(\ref{g_def}),
\[
  g(T)=\ln\lambda\left(\int\!d{\bf k}\,
  a^*_{\bf k}a_{\bf k}\e^{\omega_kT}\right)
  =T+{d\over2}\ln(2\pi T)+\ln16
\]
so eq.(\ref{Teps}) becomes
\[
  1+{d\over2T}=1+\epsilon
\]
whose solution is obviously
\begin{equation}
  T(\epsilon)={d\over2\epsilon}
  \label{page17}
\end{equation}
Now substituting $T(\epsilon)$ and $g(T)$ into eq.(\ref{fefin}) one
finds finally
\[
  f(\epsilon)={d\over2}-{d\over2}\ln{\pi d\over\epsilon}-\ln16
\]
which is nothing but the leading terms in eq.(\ref{fepsilon}).

Before considering the $O(\epsilon)$ correction to $f(\epsilon)$, let
us discuss the form of the surface of singularities $\tau=\tau_0({\bf
x})$.  From eqs.(\ref{btau}), (\ref{a}) one finds
\begin{equation}
  \tau_0({\bf x})={{\bf x}^2\over T}={2\epsilon\over d}{\bf x}^2
  \label{surface}
\end{equation}
One sees that the surface of singularity has the form of a paraboloid,
whose curvature is proportional to the typical momentum of the final
particles $k_0$.  At small $\epsilon$ the curvature radius is much
larger than the inverse boson mass, i.e. 1.  One should keep in mind,
however, that eq.(\ref{surface}) is valid only for not very large
values of ${\bf x}$ (not much larger than $k^{-1}$.  At larger ${\bf
x}$ the actual behavior of the singularity surface is unknown.

\subsubsection{Next--to--leading order}

To find the first correction to the solution, eq.(\ref{leading}), let
us substitute the ansatz
\[
  \phi_1=\sqrt{2\over\lambda}{1\over\sinh(\tau-\tau_0({\bf x}))}
  +\tilde{\phi}_1(\tau,{\bf x})
\]
to the field equation.  We expect that $\tilde{\phi_1}$ is suppressed,
compared with the leading order, by a factor of $k_0^2$, where $k_0$
is the typical momentum of the final particle, so we drop terms
containing $\phi_1^2$ and $\phi_1^3$.  We expect also that each
derivative with respect to ${\bf x}$ adds an additional factor of
$k_0$.  Denoting $y=\tau-\tau_0$, the equation for $\tilde{\phi_1}$ is
linear,
\begin{equation}
  \hat{O}\tilde{\phi}\equiv\left[\partial_\tau^2+1+
  {6\over\sinh^2y}\right]\tilde{\phi_1}
  =\sqrt{2\over\lambda}
  -\left(\partial_{\bf x}^2\tau_0\cdot{\cosh y\over\sinh^2y}+
  (\partial_{\bf x}\tau_0)^2\left({1\over\sinh y}+{2\over\sinh^3y}
  \right)\right)
  \label{Ophi1}
\end{equation}
We have ignored, for example, $\partial_{\bf x}\tilde{\phi_1}$, since
it is of order $O(k_0^4)$.  The simplest way to solve
eq.(\ref{Ophi1}) is to make use of the following simple relations,
\[
  \hat{O}\cosh y=-6{\cosh y\over\sinh^2y},\qquad
  \hat{O}\sinh y=-{6\over\sinh y}
\]
\[
  \hat{O}{1\over\sinh y}=-{4\over\sinh^3y}
\]
to see that one solution to (\ref{Ophi1}) is
\begin{equation}
  \sqrt{2\over\lambda}\left(
  {\partial_{\bf x}^2\tau_0\over6}\cosh y+{(\partial_{\bf x}\tau_0)^2\over6}
  \sinh y+{(\partial_{\bf x}\tau_0)^2\over2\sinh y}\right)
  \label{nonhom}
\end{equation}
However, one can also add to (\ref{nonhom}) an arbitrary solution to
the homogeneous equation $\hat{O}\phi=0$.  The general solution to the
homogeneous equation depends on two arbitrary functions of spatial
coordinates, $F_{1,2}({\bf x})$ and has the form,
\[
  \phi=F_1({\bf x})f_1(\tau,{\bf x})+F_2({\bf x})f_2(\tau,{\bf x})
\]
where
\[
  f_1(\tau,{\bf x})={\cosh y\over\sinh^2y}
\]
\begin{equation}
  f_2(\tau,{\bf x})=y{\cosh y\over\sinh^2y}-{\sinh y\over3}-
  {1\over\sinh y}
  \label{f2}
\end{equation}
Note that $f_1$ has a double pole at $y=\tau-\tau({\bf x})=0$, while
$f_2$ is regular on the singularity surface.  As we expect that the field
has a pole of the first order at $y=0$, so we should rule out
$f_1$ and set $F_1({\bf x})=0$ \footnote{Actually, the term
proportional to $f_1$ arises when one shifts the singularity surface
$\tau_0\to\tau_0+\delta\tau_0$ and try to expand the function
$\sinh^{-1}(\tau-\tau_0-\delta\tau_0)$ on $\delta\tau_0$.  Since we
want the singularity surface to be $\tau=\tau_0({\bf x})$, this term
should be excluded.}.  The function $F_2({\bf x})$ can be chosen from
the requirement that $\phi_1$ decreases in the Euclidean asymptotics
$\tau\to+\infty$.  In fact, in this limit (\ref{nonhom}) becomes
\[
  \sqrt{2\over\lambda}\cdot{1\over12}
  \left(\partial_{\bf x}^2\tau_0+(\partial_{\bf x}\tau_0)^2
  \right)\e^{\tau-\tau_0}
\]
while the asymptotics of $f_2$ is also growing
\[
  f_2(\tau,{\bf x})\sim-{1\over6}\e^{\tau-\tau_0}
\]
but they cancel each other when we choose $F_2({\bf
x})={1\over2}\sqrt{2\over\lambda} [\partial^2_{\bf
x}\tau_0+(\partial_{\bf x}\tau_0)^2]$.  Therefore, the solution to the
field equation with singularities at $\tau=\tau_0({\bf x})$ and decays
at $\tau\to+\infty$, up to the order of $k_0^2$, is
\[
  \phi_1=\sqrt{2\over\lambda}\left[{1\over\sinh y}+
  {\partial^2_{\bf x}\tau_0\over6}\cosh y+
  {(\partial_{\bf x}\tau_0)^2\over6}\sinh y+
  {(\partial_{\bf x}\tau_0)^2\over2\sinh y}+\right.
\]
\begin{equation}
  +\left.{\partial_{\bf x}^2\tau_0+(\partial_{\bf x}\tau_0)^2\over2}
  \left(y{\cosh y\over\sinh^2y}-
  {1\over3}\sinh y-{1\over\sinh y}\right)\right]
  \label{2order}
\end{equation}
To find $a_{\bf k}$, in complete analogy with the leading order, one
calculates the asymptotics of $\phi_1$ from eq.(\ref{2order}),
\begin{equation}
  \phi_1(\tau,{\bf x})=\sqrt{2\over\lambda}\left[\left(2-{5\over6}
  \partial_{\bf x}^2\tau_0\right)\e^{-y}+
  \left(\partial_{\bf x}^2\tau_0+(\partial_{\bf x}\tau_0)^2\right)
  y\e^{-y}\right]
  \label{2asympt}
\end{equation}
Let us note that, in contrast with the leading order, there is a term
proportional to $y\e^{-y}$ in the asymptotics of $\phi_1$.  This
structure also emerges when one evaluate the r.h.s. of
eq.(\ref{equal'}) to the next--to--leading order, replacing
$\e^{\omega_k\tau}$ by $\e^{-\tau}(1-{{\bf k}^2\over2}\tau)$.
Introducing the function
\[
  a({\bf x})=\int\!{d{\bf k}\over(2\pi)^{d/2}\sqrt{2\omega_{\bf k}}}
  \,a_{\bf k}\e^{i{\bf kx}}
\]
eq.(\ref{equal'}) reads
\begin{equation}
  \phi_1(\tau,{\bf x})=\int\!{d{\bf k}\over(2\pi)^{d/2}
  \sqrt{2\omega_{\bf k}}}\left(1-{{\bf k}^2\over2}\right)a_{\bf k}
  \e^{-\tau+i{\bf kx}}=\left(a({\bf x})+
  {\tau\over2}\partial_{\bf x}^2a({\bf x})\right)\e^{-\tau}
  \label{2asympt'}
\end{equation}
Comparing eqs.(\ref{2asympt}) and (\ref{2asympt'}), one finds the
relation between $a({\bf x})$ and $\tau(x)$, the
next--to--leading--order version of eq.(\ref{btau}),
\begin{equation}
  a({\bf x})=\sqrt{2\over\lambda}\left(2-{5\over6}\partial_{\bf x}^2
  \tau_0-\tau_0(\partial_{\bf x}^2\tau_0+(\partial_{\bf x}\tau_0)^2)
  \right)\e^{\tau_0({\bf x})}
  \label{ax}
\end{equation}
Let us find out the constraint on $a({\bf x})$ comes from the
requirement that $\tau_0(0)=0$.  Taking ${\bf x}=0$ and noticing that
$\partial_{\bf x}\tau_0|_{{\bf x}=0}=0$ since the singularity surface
is tangent to the plane $\tau=0$ at ${\bf x}=0$, it is easy to show
that this constraint is
\[
  a({\bf x})+{5\over12}\partial_{\bf x}^2a({\bf x})|_{{\bf x}=0}=
  \sqrt{8\over\lambda}
\]
In momentum representation this relation reads,
\begin{equation}
  \int\!{d{\bf k}\over(2\pi)^{d/2}\sqrt{2\omega_{\bf k}}}
  \left(1-{5\over12}{\bf k}^2\right)a_{\bf k}=\sqrt{8\over\lambda}
  \label{amom}
\end{equation}
The difference of the constraint from that of the leading order is
there is a small correction ${5\over12}{\bf k}^2$.  The calculation is
now straightforward and similar to that of the leading order.
Introducing the Lagrange multiplier and taking variation with respect
to $a_{\bf k}$, one finds
\[
  a_{\bf k}={C\over\sqrt{2\omega_{\bf k}}}\left(1-{5\over12}{\bf k}^2
  \right)
\]
where $C$ is a constant that can be found from (\ref{amom}).  The
final result of the calculations is
\[
  f(\epsilon)=
  {d\over2}\left(\ln{\epsilon\over\pi d}+1\right)+\ln16
  +\left({d-2\over4}-{5\over3}\right)\cdot\epsilon
\]
which, as expected, coincides with the perturbative result for the
tree cross section near threshold.  One can, in principle proceed
further in this direction and calculate more terms in the expansion of
$f$ on the small parameter $\epsilon$ by the same technique.  In this
way one could find $O(\epsilon^2)$ and higher corrections to
eq.(\ref{fepsilon}), which have not been found by standard
perturbative techniques.  However, let us stop here and turn to the
opposite limit $\epsilon\gg1$.

\subsection{Ultra--relativistic limit in (3+1) dimensions.  Lower
bound on tree cross section}

In the ultra--relativistic limit $\epsilon\to\infty$, the boson mass
can be neglected.  We will consider the most interesting case $d=3$,
where $\lambda$ is dimensionless.  In this case the theory becomes
scale invariant and the exponent of the cross section should be
independent of the energy (the dependence of the pre--exponent on the
energy can be determined solely by dimensional analysis).  In other
words, we expect that the function $F(\lambda n,\epsilon)$ becomes
independent of $\epsilon$ at large $\epsilon$,
\[
  \lim_{\epsilon\to\infty}F(\lambda n,\epsilon)=F(\lambda n)
\]
At tree level, this statement means that $f(\epsilon)$ has a limit
when $\epsilon\to\infty$.

{}From the first identity in eq.(\ref{dW/dn}) one sees that $T$ should
tends to 0 faster than $1/\epsilon$ (otherwise $W=\int\!dE\,T(E)$
diverges in the ultraviolet).  Now let us take the limit
$\epsilon\to\infty$ in eq.(\ref{fefin}).  One finds,
\[
  f(\infty)=-g(0)
\]
Recall that $\lambda^{-1}\e^{g(0)}$ is the minimal value of
$\int\!d{\bf k}\,a_{\bf k}^*a_{\bf k}$ over all solutions of the
massless theory that are singular at $\tau={\bf x}=0$.  We are unable
to find $g(0)$.  Instead, we will try to bound $g(0)$ from above.  We
will take a solution that is known analytically, calculate for it
$a_{\bf k}$ and plug the result into the definition of $g(0)$.  If the
extremum over $a_{\bf k}$ is the true minimum (recall our discussion
on the natrue of the extremum in subsection 3.1), this procedure would
gives us an upper bound on $g(0)$.

We take the following trial singular solution to the massless field
equation \cite{Khlebnikov},
\[
  \phi(\tau,{\bf x})=\sqrt{8\over\lambda}\,{\rho\over
  {\bf x}^2+(\tau-\rho)^2-\rho^2}
\]
The Fourier components of this configuration is
\[
  a_{\bf k}=\sqrt{8\over\lambda}\sqrt{\pi\over\omega_{\bf k}}\rho
  \e^{-\omega_k\rho}
\]
which implies the following bound on $g(0)$,
\[
  g(0)<\ln(8\pi^2)
\]
The corresponding lower bound on the cross section is
\[
  \sigma_n\geq n!\left({\lambda\over 8\pi^2}\right)^n
\]
Note that this lower bound grows factorially as $n\to\infty$, which
reflects the fact that at tree level not only amplitudes, but also
cross sections becomes large at $n\sim\lambda^{-1}$.

\section{Loop corrections at threshold}

\subsection{General consideration}

In this Section we consider another limiting case.  Namely, we take
arbitrary $\lambda n$ but small $\epsilon\ll1$, and will be interested
in the exponent of the cross section to the leading order of
$\epsilon$.  In this limit, we expect from eq.(\ref{Fpert}) that the
result has the following form,
\begin{equation}
  F(\lambda n,\epsilon)=\lambda n\ln{\lambda n\over16}-\lambda n+
  {d\over2}\left(\ln{\epsilon\over\pi d}+1\right)\lambda n
  +g(\lambda n)
  \label{gdef}
\end{equation}
where the first term in the expansion of $g(\lambda n)$ at small
$\lambda n$ should be $2\mbox{Re}B\lambda^2n^2$.  The meaning of the
function $g(\lambda n)$ becomes clear when one recalls that in the
limit $\epsilon\to0$ the cross section is the product of the square of
the threshold amplitude and the phase volume.  Dividing the cross
section corresponding to eq.(\ref{gdef}) to the phase volume at small
$\epsilon$, one finds the absolute value of the threshold amplitude,
\[
  |A_n|\propto n!\left({\lambda\over8}\right)^{n/2}
  \e^{{1\over\lambda}g(\lambda n)}
\]
Comparing with eq.(\ref{Atree}), one finds that $g$ is nothing but the
contributions of the loops to the amplitude at threshold.  Note that
the reproduction of of even the leading term in $g$,
$2\mbox{Re}B\lambda^2n^2$, is a nontrivial argument in favor to the
validity of the whole semiclassical ideology, since it shows that the
exponentiated part of loop contributions is semiclassically
calculable.

To find the form of the singularity surface, let us begin by recalling
the result of Sect.3 that $\epsilon\to0$ corresponds to $T\to\infty$
(eq.(\ref{page17})), which means that the surface of singularities has
a very large curvature radius and in the limit $\epsilon\to0$ can be
considered as a plane.  However, the discussions in Sect.3 is based on
the assumption that the limit $\lambda n\to0$ is taken.  When one
drops this assumption, one could expect that the presence of the
source at $x=0$ deforms the surface of singularities near its
location.  This change should be local and the curvature of the
singularity surface should tend to 0 at large ${\bf x}$.  At finite
$\lambda n$, one expects that the surface of singularities
$\tau=\tau_0({\bf x})$ has the form similar to that sketched in
fig.\ref{fig1}.  The requirement of zero curvature at large distances
from ${\bf x}=0$ can be satisfied if $\tau_0({\bf x})$ tends to some
constant $\tau_\infty$ as ${\bf x}\to\infty$.

If the singularity surface is just the plane $\tau=\tau_\infty$, the
solution would be ${\bf x}$--independent and equal to
\[
  \phi=\sqrt{2\over\lambda}{1\over\sinh(\tau-\tau_\infty)}
\]
In the case when the singularity surface has the form shown in
fig.\ref{fig1}, the general form of $\phi$ is
\begin{equation}
  \phi=\sqrt{2\over\lambda}\,{1\over\sinh(\tau-\tau_{\infty})}+
  \tilde{\phi}(\tau,{\bf x})
  \label{tildephi}
\end{equation}
where $\tilde{\phi}(\tau,{\bf x})$ vanishes at ${\bf x}\to\infty$ and
decays into plane waves in the asymptotics regions, since it comes
from the local (about ${\bf x}=0$) deviation of the singularity
surface from the plane $\tau=\tau_\infty$.

Let us consider the asymptotics of $\phi$ at $t\to+\infty$.  Strictly
speaking, the ${\bf x}$--independent part of $\phi$ remains
non--linear at any $t$ and does not decays into plane waves.  This is
an artifact of the limit $T\to\infty$: at any finite $T$ the field
$\phi$ is a linear superposition of waves at sufficiently large $t$.
To make $\phi$ linear, let us take the final part of the contour to
form at a small angle with the Minkowskian time axis.  Namely, we take
$t=(1+i\delta)\alpha$, where $\delta$ is small and $\alpha$ is a real
parameter going to infinity.  This leads to the replacement
\[
  {1\over\sinh(\tau-\tau_\infty)}\to2\e^{(\tau-\tau_\infty)}
\]
in the asymptotic region.  This part of $\phi$ is homogeneous on ${\bf
x}$, so its Fourier transformation is a delta--function in the
momentum space.  In contrast, the Fourier components of $\tilde{\phi}$
in the asymptotics $t\to+\infty$ are supposed to be well behaved
functions of ${\bf k}$.  Denoting $f_{\bf k}$ and $g_{\bf k}$ to be
the Feynman and anti--Feynman components of $\tilde{\phi}$ at the
final asymptotics,
\[
  \tilde{\phi}(t,{\bf k})={1\over\sqrt{2\omega_{\bf k}}}\left(
  f_{\bf k}\e^{-i\omega_kt}+g_{\bf -k}\e^{i\omega_kt}\right)
\]
one finds that the asymptotics of $\phi$ is
\begin{equation}
  \phi(t,{\bf k})={1\over\sqrt{2\omega_{\bf k}}}
   \left[f_{\bf k}\e^{-i\omega_kt}+\left(
  (2\pi)^{d/2}\sqrt{16\over\lambda}\delta({\bf k})\e^{\tau_\infty}
  +g_{\bf -k}\e^{i\omega_kt}\right)\right]
  \label{fg}
\end{equation}
Comparing eqs.(\ref{phi_bc}) and (\ref{fg}), one finds,
\begin{equation}
  b_{\bf k}\e^{\omega_kT-\theta}=f_{\bf k}
  \label{bf}
\end{equation}
\begin{equation}
  b_{\bf k}^*=(2\pi)^{d/2}\sqrt{16\over\lambda}\delta({\bf k})
  \e^{\tau_\infty}+g_{\bf k}
  \label{bg}
\end{equation}
{}From eq.(\ref{bf}) one obtains $b_{\bf k}=f_{\bf
k}\e^{-\omega_kT+\theta}$.  Since $T$ is large, $b_{\bf k}$ looks like
a delta--function in the ${\bf k}$--space: $b_{\bf k}$ is non--zero
only in a small region near ${\bf k}=0$, namely, $k\sim T^{-1/2}$.  As
$f_{\bf k}$ is a smooth function of ${\bf k}$, one can replace $f_{\bf
k}$ by its value at ${\bf k}=0$ and write
\[
  b_{\bf k}=f_0\e^{-\omega_kT+\theta}
\]
Consider now eq.(\ref{bg}).  Since $b^*_{\bf k}$ has the form of the
delta--function, $g_{\bf k}$ vanishes.  This important result means
that $\tilde{\phi}(t,{\bf x})$ is purely Feynman in the final
asymptotics.  Eq.(\ref{bg}) now reads
\begin{equation}
  f_0^*\e^{-\omega_kT+\theta}=(2\pi)^{d/2}\sqrt{16\over\lambda}
  \delta({\bf k})\e^{\tau_\infty}
  \label{f0T}
\end{equation}
This equation should be understood as a symbolic representation of a
relation between $f_0$, $T$, $\theta$ and $\tau_\infty$.  To find the
latter let us take the integral of both sides of eq.(\ref{f0T}) over
$d{\bf k}$.  We obtain
\[
  f_0^*=\sqrt{16\over\lambda}\,T^{d/2}\e^{T-\theta+\tau_{\infty}}
\]
The multiplicity and the (small) kinetic energy of the final state can
be also expressed in terms of $T$, $\theta$ and $\tau_\infty$,
\begin{equation}
  n=\int\!d{\bf k}\,b^*_{\bf k}b_{\bf k}\e^{\omega_kT-\theta}
  ={16\over\lambda}(2\pi T)^{d/2}\e^{T-\theta+2\tau_{\infty}}
  \label{n}
\end{equation}
\begin{equation}
  n\epsilon=\int\!d{\bf k}\,{{\bf k}^2\over2}b^*_{\bf k}b_{\bf k}
  \e^{\omega_kT-\theta}=
  {16\over\lambda}(2\pi T)^{d/2}\e^{T-\theta+2\tau_{\infty}}
  \cdot{d\over2T}
  \label{E}
\end{equation}
Solving eqs.(\ref{n}) and (\ref{E}) with respect to $T$ and $\theta$,
one finds,
\[
  T={d\over2\epsilon}
\]
\[
  \theta=-\ln{\lambda n\over16}+{d\over2}\left(\ln{\pi d\over\epsilon}
  +1\right)+2\tau_{\infty}
\]
Note that $T$ and $\theta$ depend on the surface of singularities
through $\tau_\infty$ but are insensitive of the behavior of the
function $\tau_0({\bf x})$ at finite ${\bf x}$.  In contrast, the
action depends on the the precise form of this surface.  In Sect.3 we
have seen that a part of the action (that is linear on $\tilde{\phi}$)
can be reduced to the boundary term
$\tilde{\phi}\partial_t\phi|_{t=\infty}$ and is equal to ${in\over2}$.
It is convenient to separate this contribution from the action and
introduce the notation $S'=\mbox{Im}S-{n\over2}$.  The exponent of the
cross section is now equal to
\begin{equation}
  W=ET-n\theta-2\mbox{Im}S=
  {dn\over2}\left(ln{\pi d\over\epsilon}+1\right)+
  n\ln{\lambda n\over16}-n-2S'[\tau_0({\bf x})]-2n\tau_{\infty}
  \label{WS'}
\end{equation}
where $S'$ a functional depending on $\tau_0({\bf x})$.  Notice that
the first three terms in eq.(\ref{WS'}) coincide with those in
eq.(\ref{gdef}), we conclude that the exponent for the loop corrections
to the amplitude at threshold can be obtained by maximizing the
expression
\begin{equation}
  {1\over\lambda}g(\lambda n)=-n\tau_\infty-S'[\tau_0({\bf x})]
  \label{gS'}
\end{equation}
with respect to all possible surfaces of singularities $\tau_0({\bf
x})$.  The problem has a simple geometric interpretation: it is
equivalent to finding the equilibrium configuration of
a surface under the force $n$ acting to the point ${\bf x}=0$, when
the energy of the surface depends on its form through the functional
$S'[\tau_0({\bf x})]$.

Let us summarize our results.  To find the amplitude at threshold, one
should
\begin{itemize}
\item[1.] Find for any function $\tau=\tau_0({\bf x})$ (which goes to
a constant value $\tau_\infty$ as ${\bf x}\to\infty$) the solution to
the field equation which is singular on the surface $\tau=\tau_0({\bf
x})$ and decays at $\tau\to+\infty$ and at $t=+\infty$
has the form (\ref{tildephi}) where $\tilde{\phi}$ is purely Feynman.
\item[2.] calculate the action of the field, and find
$S'=\mbox{Im}S-{n\over2}$
\item[3.] Extremize the r.h.s. of eq.(\ref{gS'}) over all surfaces
$\tau=\tau_0({\bf x})$.
\end{itemize}

Unfortunately, $S'$ seems to be a very complicated functional of
$\tau_0({\bf x})$.  One region when it can be calculated is the regime
$\tau_0\ll1$, where one can develop perturbation theory on $\tau_0$.
Let us see that this is the regime of low multiplicities,
$n\ll1/\lambda$.

\subsection{Low multiplicities, $\lambda n\ll1$}

When $\tau_0({\bf x})$ is small the field configuration can be found
by solving the field equation separately at $\tau\ll1$ and
$\tau\gg\tau_{\infty}$ and matching in the intermediate region
$\tau_0\ll\tau\ll1$.  For convenience we will deform the contour as
shown in fig.\ref{fig2}, where $\tau_c$ is some value in the
intermediate region, $\tau_{\infty}\ll\tau_c\ll1$. So, the contour in
the $t$ plane consists of four parts: (I) $(i\infty,i\tau_c)$, (II)
$(i\tau_c,0)$, (III) $(0,i\tau_c)$, (IV) $(i\tau_c,i\tau_c+\infty)$.
The energy on parts (I) and (II) is 0, while on parts (III) and (IV)
of the contour it is equal to $E$.

First consider parts (II) and (III).  In these part $\tau\ll1$, and
the field is given by the following formula,
\begin{equation}
  \phi(\tau,{\bf x})=\sqrt{2\over\lambda}\,{1\over\tau-\tau_0({\bf
x})}+\mbox{small corrections}
  \label{smalltau}
\end{equation}
Eq.(\ref{smalltau}) is valid on both parts (II) and (III).  While the
leading singular terms coincide, there is difference between $\phi$ on
(II) and (III).  Denote
\[
  \delta\phi(\tau,{\bf x})=\phi_{III}(\tau,{\bf x})-
  \phi_{II}(\tau,{\bf x})
\]
Since we expect that this difference is much smaller than the leading
term in eq.(\ref{smalltau}), $\delta\phi$ satisfies the linearized
equation
\[
  (\partial_{\tau}^2+\partial_{\bf x}^2+1+3\lambda\phi_0^2)
  \delta\phi=0
\]
Furthermore, at $\tau\ll1$, the mass and the spatial derivatives can
be neglected, so $\delta\phi$ satisfies the equation
$(\partial_{\tau}^2+3\lambda\phi_0^2)\delta\phi=0$, the general
solution to which that vanishes on the singularity surface is
\begin{equation}
  \delta\phi(\tau,{\bf x})=\sqrt{2\over\lambda}\,
  W({\bf x})(\tau-\tau_0({\bf x}))^3
  \label{deltaphi}
\end{equation}
where the function $W({\bf x})$ is yet to be determined.

To the linear order on $\delta\phi$, the action on the parts (II) and
(III) can be reduced to the boundary term at $\tau=\tau_c$,
\begin{equation}
  S'_{II+III}=-\int\!d{\bf x}\,
  \partial_\tau\phi\cdot\delta\phi|_{\tau=\tau_c}=
  {2\over\lambda}\left((\tau_c-\tau_\infty)
  \int\!d{\bf x}\,W({\bf x})-
  \int\!d{\bf x}\,W({\bf x})c({\bf x})\right)
  \label{Ssmalltau}
\end{equation}
where $c({\bf x})=\tau_0({\bf x})-\tau_\infty$.  The next correction
to $S'_{II+III}$ is suppressed by a factor of $\tau_\infty^2$ compared
to the leading result (\ref{Ssmalltau}) and will be neglected.

Consider now the parts (I) and (IV). On these parts of the contours,
the solutions to the field equation can be represented in the form of
the perturbative series on the background of the homogeneous field,
\begin{equation}
  \phi=\phi_{(0)}+\phi_{(1)}+\phi_{(2)}+\cdots
  \label{largetau}
\end{equation}
where $\phi_{(0)}$ is the ${\bf x}$--independent solution to the field
equation,
\begin{equation}
  \phi_0=\sqrt{2\over\lambda}\,{1\over\sinh(\tau-\tau_\infty)}
  \label{phi0}
\end{equation}
$\phi_{(1)}$ is the linear wave on the background (\ref{phi0})
satisfying the equation
\begin{equation}
  \partial_\mu^2\phi_{(1)}-V'(\phi_{(0)})\phi_{(1)}=0
  \label{phi(1)}
\end{equation}
everywhere except the singularities and having the following explicit
form,
\begin{equation}
  \phi_{(1)}=\sqrt{2\over\lambda}
  \int\!{d{\bf k}\over(2\pi)^d}\,{1\over3}c_{\bf k}f^k_2(\tau)
  \e^{i{\bf kx}},\qquad\mbox{on part (I)}
  \label{phi11}
\end{equation}
\begin{equation}
  \phi_{(1)}=\sqrt{2\over\lambda}
  \int\!{d{\bf k}\over(2\pi)^d}\,{1\over3}c_{\bf k}f^k_1(\tau)
  \e^{i{\bf kx}},\qquad\mbox{on part (IV)}
  \label{phi14}
\end{equation}
where $c_{\bf k}$ are some function of ${\bf k}$, and
$f_{1,2}^k(\tau)$ are mode functions on the background field
(\ref{phi0}) \cite{Volprop,AKPprop},
\[
  f^k_1(\tau)=\e^{\omega_k(\tau-\tau_\infty)}\left(\omega_{\bf k}^2-
  {3\omega_{\bf k}\over\tanh(\tau-\tau_\infty)}+2+
  {3\over\sinh^2(\tau-\tau_\infty)}\right)
\]
\[
  f^k_2(\tau)=f^k_1(2\tau_\infty-\tau)
\]
and $\phi_{(2)}$, etc. are higher corrections.  The small parameter
governing the expansion (\ref{largetau}) is in fact $\tau_\infty$.

The relation between $c_{\bf k}$ and the form of the surface of
singularities can be found from the matching condition between
eqs.(\ref{largetau}) and (\ref{smalltau}) at intermediate values of
$\tau$.  One finds,
\[
  c_{\bf k}=\int\!d{\bf x}\,c({\bf x})\e^{i{\bf kx}}
\]
One can also relate $c_{\bf k}$ with the function $W({\bf x})$.  In
fact, at $\tau=\tau_c\ll1$ one finds from eqs.(\ref{phi11}) and
(\ref{phi14})
\begin{equation}
  \delta\phi_{(1)}=\sqrt{2\over\lambda}{(\tau_c-\tau_\infty)^3\over45}
  \int\!{d{\bf k}\over(2\pi)^d}\,W_{\bf k}c_{\bf k}\e^{i{\bf kx}}
  \label{deltaphi(1)}
\end{equation}
where
\begin{equation}
  W_{\bf k}=2\omega_{\bf k}(\omega^2_{\bf k}-1)(\omega^2_{\bf k}-4).
  \label{Wk}
\end{equation}
On the other hand, since $\tau_c\gg\tau_0$ eq.(\ref{deltaphi}) can be
expanded on the parameter $\tau_0/\tau_c$ as follows,
\begin{equation}
  \delta\phi=\sqrt{2\over\lambda}W({\bf x})(\tau_c-\tau_\infty)^3
  -3W({\bf x})c({\bf x})(\tau_c-\tau_\infty)^2+\cdots
  \label{deltaphi(1)'}
\end{equation}
The r.h.s. of eq.(\ref{deltaphi(1)}) should coincide with the first
term of the r.h.s. of (\ref{deltaphi(1)'}), so we obtain the following
relation,
\[
  W({\bf x})={1\over45}
  \int\!{d{\bf k}\over(2\pi)^d}\,W_{\bf k}c_{\bf k}\e^{i{\bf kx}}
\]
Let us now evaluate the action on the part (I).  Up to the second
order of the small parameter the action has the form
\[
  S_I'=S_I[\phi_0]+\int\limits_{\tau_c}^\infty\!\left(
  \partial_\mu\phi_{(0)}\partial_\mu\phi_{(1)}+
  \partial_\mu\phi_{(0)}\partial_\mu\phi_{(2)}+
  {1\over2}(\partial_\mu\phi_{(1)})^2+
  V'(\phi_{(0)})(\phi_{(1)}+\phi_{(2)})+\right.
\]
\begin{equation}
  \left.{1\over2}V''(\phi_{0})(\phi_{(1)})^2\right)
  \label{SI'}
\end{equation}
Making use of eq.(\ref{phi(1)}) and the fact that $\phi_{(0)}$
satisfies the field equation, the integral in eq.(\ref{SI'}) can be
taken in part and the result is
\[
  S_I'=S_I[\phi_0]-\int\!d{\bf x}\left(
  (\phi_{(1)}+\phi_{(2)})\partial_\tau\phi_{(0)}+
  {1\over2}\phi_{(1)}\partial_\tau\phi_{(1)}
  \right)|_{\tau=\tau_c}
\]
The action on the part (IV) can be calculated in an analogous way.
Since $S_I[\phi_0]+S_{IV}[\phi_0]=0$, the sum of the action on parts
(I) and (IV) is given by the following boundary terms,
\begin{equation}
  S'_{I+IV}=\int\!{d{\bf x}}\,
  \left(\partial_\tau\phi_{(0)}\cdot\delta\phi_{(1)}+
  \partial_\tau\phi_{(0)}\cdot\delta\phi_{(2)}+
  {1\over2}\partial_\tau\phi_{(1)}\cdot\delta\phi_{(1)}+
  {1\over2}\partial_\tau\delta\phi_{(1)}\cdot\phi_{(1)}\right)_{\tau=\tau_c}
  \label{S14}
\end{equation}
The analytical expressions for $\phi_2$ are rather complicated.
Fortunately, the calculation of the action requires only
$\delta\phi_2$ at $\tau=\tau_c$, which is equal to the second term in
the r.h.s. of eq.(\ref{deltaphi(1)'}),
\[
  \delta\phi_2(\tau_c,{\bf x})=-3W({\bf x})c({\bf x})(\tau_c-\tau_\infty)^2
\]
Substitute this to eq.(\ref{S14}) one obtains
\begin{equation}
  S'_{I+IV}={2\over\lambda}\left(-(\tau_c-\tau_\infty)
   \int\!d{\bf x}\,W({\bf x})+
   {7\over2}\int\!d{\bf x}\,W({\bf x})c({\bf x})\right)
  \label{Slargetau}
\end{equation}
The full action can be obtained by taking the sum of (\ref{Slargetau})
and (\ref{Ssmalltau}). The dependence on $\tau_c$ disappears, as one
can anticipate, and the action is quadratic on $c_{\bf k}$,
\begin{equation}
  S'={5\over\lambda}\int\!d{\bf x}\,W({\bf x})c({\bf x})=
  {1\over9\lambda}\int\!{d{\bf k}\over(2\pi)^d}\,
  W_{\bf k}|c_{\bf k}|^2
  \label{56*}
\end{equation}
Now to extremize (\ref{gS'}) we note that
\begin{equation}
  \tau_\infty=-c(0)=\int\!{d{\bf k}\over(2\pi)^d}\,c_{\bf k}
  \label{**}
\end{equation}
so
\begin{equation}
  {1\over\lambda}g=n\int\!{d{\bf k}\over(2\pi)^d}\,c_{\bf k}-
  {1\over9\lambda}\int\!{d{\bf k}\over(2\pi)^d}\,W_{\bf k}
  |c_{\bf k}|^2
  \label{***}
\end{equation}
Differentiating $g$ with respect to $c_{\bf k}$, one obtains
\begin{equation}
  c_{\bf k}={9\lambda\over2}{n\over W_{\bf k}}
  \label{*'}
\end{equation}
and
\[
  {1\over\lambda}g={9\lambda\over4}n^2\int\!{d{\bf k}\over(2\pi)^d}\,
  {1\over W_{\bf k}}=B\lambda n^2
\]
which coincides with the perturbative result for exponentiated
leading--$n$ loops.

Let us find the condition for the approximation we use here to be
valid.  The small parameter is $\tau_\infty$, and from eqs.(\ref{**})
and (\ref{*'}) we see that $\tau_\infty\sim\lambda n$.  So, our
calculations are reliable when $n$ is small, $n\ll1/\lambda$.

Another remark should be made on the nature of the extremum.  Since
$W_{\bf k}$ does not have definite sign (it is negative at ${\bf
k}^2<3$ and positive at ${\bf k}^2>3$, see eq.(\ref{Wk})), the
extremum of the r.h.s. of (\ref{***}) is neither maximum nor minimum,
but rather a saddle point (in contrast with the maximum for
tree--level cross section considered in sect.3).  In the theory with
broken symmetry $W_{\bf k}$ is a positively defined function
\cite{Smith}, and we have in this case the true minimum.

\section{Conclusion}

We have seen that the problem of calculation multiparticle cross
sections can be reduced to a certain problem of the classical theory.
The field configuration describing these processes is a singular
solution to the field equation with certain boundary conditions that
optimizes the transition rate.  Though most results obtained in this
paper can also be found by making use of various perturbative methods,
our discussions show that they can be derived from a single approach.
We have pointed out an important fact that the exponentiated loop
corrections can be calculated semiclassically.  We also obtain a new
result, namely, the lower bound on tree cross section in the
ultra--relativistic regime.  Hopefully, the formalism developed in
this paper can be used in further analytical or numerical calculations
of the multiparticle cross section

The author would like to thank V.A.~Rubakov and P.G.~Tinyakov for
numerous stimulating discussions and valuable comments.  The author is
grateful to L.~McLerran and M.B.~Voloshin for discussions of the
results, and thanks Rutgers University and Theoretical Physics
Institute, University of Minnesota, for hospitality.  This work is
supported, in part, by Russian Foundation for Fundamental Research,
grant \# 93-02-3812 and INTAS grant \# INTAS-93-1630.

\newpage

\setlength{\unitlength}{1.5cm}
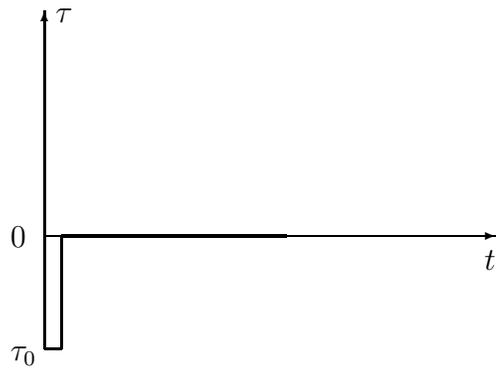
\begin{figure}
\begin{center}
\begin{picture}(4,3)
\put(0,1){\vector(1,0){4}}
\put(0,0){\vector(0,1){3}}
\thicklines
\put(0,0){\line(0,1){2}}
\put(0,0){\line(1,0){.15}}
\put(.15,0){\line(0,1){1}}
\put(.15,1){\line(1,0){2}}
\thinlines
\put(0.1,2.9){$\tau$}
\put(3.9,0.7){$t$}
\put(-.3,0.9){0}
\put(-.3,-0.1){$\tau_0$}
\end{picture}
\end{center}
\caption{The contour.}
\label{contour}
\end{figure}

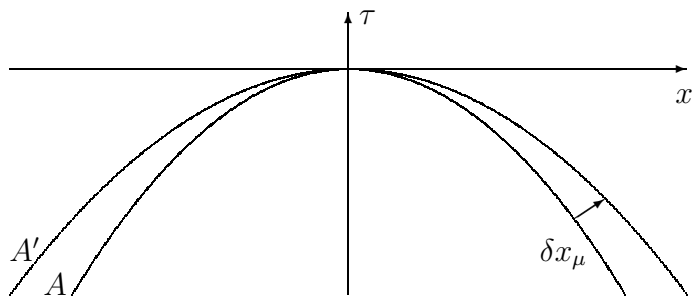
\begin{figure}
\begin{center}
\begin{picture}(6,2)(0,1)
\put(3,1){\vector(0,1){2.5}}
\put(3.1,3.4){$\tau$}
\put(5.9,2.7){$x$}
\put(0,3){\vector(1,0){6}}
\bezier{500}(0.55,1)(3,5)(5.45,1)
\bezier{500}(0,1)(3,5)(6,1)
\put(5,1.67){\vector(3,2){0.27}}
\put(4.7,1.3){$\delta x_\mu$}
\put(0.3,1){$A$}
\put(0,1.3){$A'$}
\end{picture}
\end{center}
\caption{The deformation of the singularity surface.}
\label{shift}
\end{figure}

\setlength{\unitlength}{1cm}
\begin{figure}
\begin{center}
\begin{picture}(6,3)
\put(3,0){\vector(0,1){3}}
\put(0,1){\vector(1,0){6}}
\put(0,0){\line(1,0){1}}
\put(5,0){\line(1,0){1}}
\put(3,0){\line(1,0){.1}}
\bezier{100}(1,0)(1.5,0)(2,0.5)
\bezier{100}(2,0.5)(2.5,1)(3,1)
\bezier{100}(3,1)(3.5,1)(4,0.5)
\bezier{100}(4,0.5)(4.5,0)(5,0)
\put(3.1,2.9){$\tau$}
\put(5.9,0.7){$t$}
\put(3.2,0){$\tau_\infty$}
\end{picture}
\end{center}
\caption{The surface of singularities for configurations describing
processes at exact threshold.}
\label{fig1}
\end{figure}
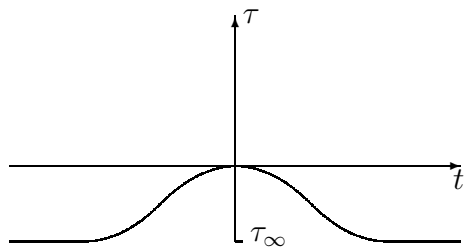

\setlength{\unitlength}{1.5cm}
\begin{figure}
\begin{center}
\begin{picture}(4,3)
\put(0,0){\vector(1,0){4}}
\put(0,0){\vector(0,1){3}}
\put(0,1){\line(-1,0){.1}}
\thicklines
\put(0,0){\line(0,1){2}}
\put(0,0){\line(1,0){.15}}
\put(.15,0){\line(0,1){1}}
\put(.15,1){\line(1,0){2}}
\thinlines
\put(0.1,2.9){$\tau$}
\put(3.9,-0.3){$t$}
\put(-.3,0.9){$\tau_c$}
\put(-.3,-0.1){$\tau_0$}
\put(-.3,1.4){I}
\put(-.3,0.4){II}
\put(0.25,0.4){III}
\put(1,1.1){IV}
\end{picture}
\end{center}
\caption{The contour for calculating the action at low multiplicities.}
\label{fig2}
\end{figure}
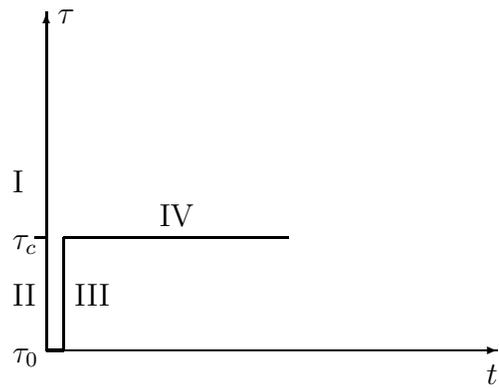

\end{document}